**A clearer approach for defining unit systems**

Paul Quincey and Richard J. C. Brown

Environment Division, National Physical Laboratory, Hampton Road, Teddington, TW11 0LW, United Kingdom.

Tel: +44 (0)20 8943 6788; E-mail address paul.quincey@npl.co.uk

Abstract

We present the SI and other unit systems, including cgs-em and cgs-es, in a framework whereby a system of fully independent and dimensionally orthogonal base units is modified by conventions designed to simplify the equations that are used within each system. We propose that the radian can be seen as an independent unit whose dimensional status is modified in the SI and other unit systems for this purpose.

This framework clarifies how different unit systems are interrelated, and identifies the key pieces of information that are needed to define both a unit system and the equations that are to be used with it. Specifically, these are the size of the base units in the unsimplified system, together with sufficient equations to identify all the conventions adopted by the particular unit system. The appropriate extra information for the revised SI is presented. We do not propose that the treatment of angles as dimensionless within the SI is changed.

It is also proposed that the Gaussian unit system is best seen as identical to cgs-es, but with the *B* and *H* symbols in equations used to represent "relativistic" versions of *B* and *H*, which should properly be treated as different quantities. These different versions of *B* and *H* can similarly be used within the SI, with many of the advantages of the Heaviside-Lorentz system.

Introduction

Systems of measurement units are, above all, intended to be useful and practical. There is no objectively correct system; the SI (BIPM, SI Brochure 8th edition, 2006) was designed to fulfil the need for a common system covering very broad areas of measurement. It can provide comparability at high accuracy, with many of the peculiarities of historical measurement systems removed, but without losing all their familiar features or convenience. Since its adoption in 1960, the SI has evolved to accommodate changing priorities, as was always intended. It is currently undergoing significant changes in the way its base units are defined, most notably replacing the definition of the kilogram in terms of a unique metal cylinder, and this change is accompanied by a revision of the SI Brochure that includes a shift in emphasis from base units to defining constants (BIPM, SI Brochure draft 9th edition, 2016). It is therefore timely to consider how the most important elements of a system of units can best be summarised.

The role of a system of units, and the example of electrical units

While measurement units have been with us throughout recorded history, systems of units have only been developed since the 19th Century. The term "system of units" is here used to mean a set of units designed to cover a large range of measureable quantities, which has some of the known scientific relationships between these quantities built-in, to reduce the number of independent[1]

---

[1] We use the term "independent units" to mean a set of units defined in such a way that the size of one of the units can be changed without altering the size of the other units in the set. The SI base units are independent in this sense, even though changing the size of one unit may require numerical values within the definitions of other units to be changed, so that they remain the same size. It is not possible to change the size of an SI

units needed. These include not just obvious relationships, like units for volume being the cube of those for length, but less obvious ones, like the unit for electrical energy being the same as those for kinetic energy and heat. One key period was 1861-1873, when the issue was addressed by committees, which included James Clerk Maxwell and William Thomson (later Lord Kelvin), within the British Association for the Advancement of Science (BA). The BA committees proposed a clear rationale for such unit systems in terms of a small number of independent base quantities, with associated base units, from which all other units are derived (de Boer, 1995; Roche, 1998).

A key decision in the process of setting up the SI concerned electrical and magnetic quantities. If charge is considered to be independent of mass, length and time, the equation describing the force between two electric charges in a vacuum must include a dimensional constant $k_C$, so that $F = k_C q_1 q_2 / r^2$, where the other symbols have their usual meanings. This was the choice made by the SI, where the equation to be used is $F = q_1 q_2 / 4\pi\varepsilon_0 r^2$, and $\varepsilon_0$ is the permittivity of free space with units $m^{-3} kg^{-1} s^4 A^2$. The decision to represent $k_C$ as $1/4\pi\varepsilon_0$ includes a choice to place a factor of $4\pi$ in this equation to avoid the need to do so in Maxwell's equations. The process of avoiding the factors of $4\pi$ in Maxwell's equations is termed rationalisation. These decisions followed the proposal of Giovanni Giorgi in 1901 (Frezza et al, 2015).

So, apart from defining the sizes of the various units to be used, a system of units also determines, at least for electromagnetism, conventions regarding which of several possible equations must be used. Both aspects are of great importance to people who will use the system, and should therefore be set out as clearly as possible.

The example of angle units

Another major decision when setting up the SI concerned the status of angles within the system. In practice this was not a decision about which set of equations was to be used – only one set was in use throughout science – but about how the status of angle units within the SI could be made compatible with this set of equations. The anomalous handling of the radian and steradian within the SI, and indeed all other unit systems, whereby plane angle and solid angle have independently defined units, radians and steradians, as if they are base quantities, but are also treated as if they are dimensionless (with the unit "m/m" or "$m^2/m^2$" respectively), has been widely discussed elsewhere (eg Brinsmade, 1936; Torrens, 1986; Brownstein, 1997; Mohr and Phillips, 2014; Quincey and Brown, 2016; Quincey, 2016). Between 1960 and 1995 the radian and steradian belonged to a specially created category of "supplementary units".

The SI decision can be summarised as requiring that the equation $s = r\,\theta$ is to be used for the relationship between the arc length to the radius of a circle, when the equation $s = \eta\, r\, \theta$, where $\eta$ is a constant with the dimension angle$^{-1}$ and value 1 rad$^{-1}$, would fit more naturally with the formalism used for the other units within the SI[2]. In the same way, the equation $A = \Omega\, r^2$ is to be used to relate the area on the surface of a sphere to the solid angle $\Omega$ at its centre, instead of $A = \eta^2\, \Omega\, r^2$. Recent proposals to treat the radian as an SI base unit (eg Mohr and Phillips, 2014) would, in effect, require the alternative, unfamiliar equations to be used, in the same way that treating electrical current as independent of mechanical quantities requires the additional constant described above.

In the proposed wording of the new SI Brochure (BIPM, draft 9$^{th}$ edition of the SI Brochure, 2016), the definition "one radian is the angle subtended at the centre of a circle by an arc that is equal in

---

derived unit, such as a joule, without either changing the size of another unit or losing the coherence of the system.

[2] We use the symbol $\eta$ for this constant, as given in Torrens (1986) and elsewhere. Brownstein (1997) used the symbol □; Brinsmade (1936) and several others did not use a special symbol, but adopted the unit symbol "rad" for use as a constant, so that $\eta = 1/rad$. The constant $\eta$ can also be represented as $1/\theta_N$, where $\theta_N = 1$ rad, the "natural" unit of angle.

length to the radius" does not in itself define the equation that relates these quantities. This can be deduced from the additional information that the radian is considered to have the unit "m/m". We think that it would be much more useful to make the required equations explicit. These equations would also provide a much more concrete focus for discussions about possible changes to how the radian and steradian are treated by the SI than an abstract focus on the status of the units.

A classification of unit systems

For the purposes of this part of the paper, the quantities included in a unit system are limited to those in mechanics and electromagnetism. Other topics such as temperature, chemistry, and photometry are omitted for reasons of simplicity. We propose that this reduced scope is covered by five independent base quantities, length, mass, time, charge and angle.

The choice of five base units is for pragmatic reasons and will not be rigorously justified. The question that has been answered is not "How many independent physical dimensions are there?" but "How many base quantities are needed to reduce the incidence of different quantities being given the same unit to an acceptable level?" There are many possibilities for the five quantities that are included, and the choice is based primarily on the choices within existing unit systems, for familiarity.

These are very similar to the base quantities in the SI, apart from angle, which needs some explanation. We present three arguments for its inclusion. Firstly, at a fundamental level, Noether's theorem links the independent conservation laws for energy, momentum and angular momentum to symmetries in time, position and orientation, suggesting that angle has a fundamental role in the "playing field" of physics along with length and time. Secondly, if we take the incidence of different quantities being ascribed the same units as an indication that we have an insufficient number of base quantities, the fact that torque and energy, and action and angular momentum, have common units suggests that an extra quantity is needed. And lastly, as has already been noted, many people have pointed out that the exclusion of angle as a base quantity appears to be anomalous. As explained in the previous section, there is a good reason why this anomaly exists within the SI, which is that it provides compatibility with the equations habitually used within mathematics and physics.

The premise of this classification of unit systems is that there is a trade-off between maximum clarity of units – different quantities having different units, allowing dimensional analysis to be most effective – and simplicity of equations – specifically the removal of constants that appear in the equations used in the unsimplified system. These constants are generally termed either "conversion factors" or "fundamental constants", the distinction between the two often being debatable (e. g. Wilczek, 2007; Duff, 2014). Understandably, theoretical physicists tend to prefer simpler equations and take the identification of the physical quantity more for granted than other users of unit systems. In this classification, all unit systems have a base unit for all five base quantities, but this may not be explicitly apparent after the various conventions leading to the simplifications have been made.

The classification therefore starts with an unsimplified system. The five base quantities have independent base units and equations are written accordingly, with maximum transparency and a maximum number of constants within equations.

As shown in Table 1, the SI does *not* form an unsimplified system, because of its treatment of angle (common to all familiar unit systems), as described in Quincey (2016). The unsimplified system is therefore called the "underlying SI". Definitions of the base units are not given here; the precise definitions of some of these have changed over time, but the quantities they represent have remained the same for most practical purposes since their creation.

It should be emphasised at this point that while an unsimplified unit system brings maximum clarity to dimensional analysis, it can do nothing to distinguish between different types of dimensionless quantity. For example, strain, refractive index, and the number of soot particles emitted by a diesel engine are utterly different quantities. Such dimensionless quantities are not identified by the unit system; their identification must come from clear verbal descriptions.

Procedures for simplifying equations

In general terms, equations in physics can be simplified - removing a constant - by choosing the measurement units to make the constant numerically equal to 1, and additionally treating the constant as dimensionless. This can be done in two related ways:

(1) The "natural base unit" procedure. Any base quantity, such as a mass, $M$, can be treated as dimensionless *within equations* by making the symbol $M$ represent the ratio $M/M_U$, where $M_U$ is the chosen base unit of mass. Any dimensional constant in the equation can be made dimensionless by making all the base quantities relating to it dimensionless in this way. The constant can be made equal to the number one by careful choice of the base units - in the equations of physics these will usually be "natural units", such as the mass of an electron, $m_e$. In that specific case the procedure can be described as "setting $m_e$ equal to 1", though in fact $m_e$ still has the same size and dimension, and it is the convention for representing mass in equations that has changed. It would clearly be wrong to consider that the simplified equation "tells us" that the mass of an electron is the same quantity as the number one, as this is just the convention we have used to simplify the equation.

(2) The "linked base units" or "natural derived unit" procedure. Any dimensional constant within an equation, such as the speed of light, $c$, can be made numerically equal to 1 by linking together the relevant base units, in this case those for length and time such that $L_U = cT_U$, so that $c = 1\ L_U/T_U$, still a dimensional quantity. To make $c$ equal to the number one, $L_U$ and $T_U$ must also be treated as having the same dimension. More generally, the process requires that one base quantity is treated as having a new, unaccustomed dimension - some combination of the other dimensions - for convenience. This "setting the speed of light equal to 1" within an equation requires that a second of time is treated as the same quantity as 299 792 458 metres. We can see how these quantities are related through special relativity, but they are not the same thing, at least not for practical purposes; again this is a convention to simplify an equation.

To take another example, setting the gravitational constant $G$ equal to 1 requires a mass of one kilogram to be considered to be the same quantity as 6.67 x $10^{-11}$ $m^3/s^2$. Again we can see how the two quantities are related, in this case in terms of the gravitational acceleration of a mass at some distance from a mass of one kilogram, but they are not the same thing. It should be no surprise that a procedure for simplifying equations, although mathematically sound, does not provide new physical insights. We therefore suggest that choosing to set $k_C$ equal to one, as in the cgs-es system, cannot be said to show that a charge of one statcoulomb is the same quantity as one $cm^{3/2} g^{1/2} s^{-1}$ in any meaningful physical sense.

The "linked base units" procedure can be turned into a "natural base unit" procedure by changing the base quantities in the unit system, specifically by choosing the dimensions of the constant that is to be removed from the equation to be those of a base quantity. The distinction between the procedures is also blurred when many units are involved, as in the "atomic" and "Planck" rows of Table 1. Each use of either procedure has the effect of removing a constant from equations and embedding it in the unit system, where it is much less visible, and also of removing a dimension from the system, making dimensional analysis less useful.

Table 1 shows the size of the five base units in a selection of unit systems, together with any simplifying conventions. Where a base unit is treated as a natural unit, the size of the natural unit is put in brackets. Where base units are linked together this is shown by the equation in the unit box.

As already mentioned, the radian is treated as a natural unit in all simplified systems. In addition, two methods of linking base units to simplify the equations of electromagnetism are used to form the cgs-em and cgs-es systems. The cgs-Heaviside-Lorentz units are a refinement of cgs-es, with the aim of rationalising the equations of electromagnetism. Two sets of "natural units" – atomic and Planck – are included to show how they fit into this framework.

| Unit system | Length L | Mass M | Time T | Charge Q | Angle A | Units for $k_C$ and angular momentum | Features |
|---|---|---|---|---|---|---|---|
| underlying SI | m | kg | s | C | rad | $m^3 \cdot kg \cdot s^{-2} \cdot C^{-2}$<br>$m^2 \cdot kg \cdot s^{-1} \cdot rad^{-1}$ | Note 1 |
| SI | m | kg | s | C | (1 rad) | $m^3 \cdot kg \cdot s^{-2} \cdot C^{-2}$<br>$m^2 \cdot kg \cdot s^{-1}$ | $\eta = 1$<br>Note 1 |
| cgs-em | cm | g | s | abC ≈ 10.0 C (Note 2)<br>1 abC = 1 $cm^{1/2} g^{1/2}$ | (1 rad) | $cm^2 \cdot s^{-2}$<br>$cm^2 \cdot g \cdot s^{-1}$ | $\eta = 1$<br>$k_C = c^2$ |
| cgs-es | cm | g | s | statC ≈ 3.34 x $10^{-10}$ C<br>(Note 3)<br>1 statC = 1 $cm^{3/2} g^{1/2} s^{-1}$ | (1 rad) | 1<br>$cm^2 \cdot g \cdot s^{-1}$ | $\eta = k_C = 1$ |
| cgs-Heaviside-Lorentz | cm | g | s | "HL"C ≈ 9.41 x $10^{-11}$ C<br>1 "HL"C = 1 statC/$2\pi^{1/2}$<br>= 1 $cm^{3/2} g^{1/2} s^{-1}/2\pi^{1/2}$ | (1 rad) | 1<br>$cm^2 \cdot g \cdot s^{-1}$ | $\eta = 1$<br>$k_C = 1/4\pi$ |
| atomic | (5.29 x $10^{-11}$ m) | (9.11 x $10^{-31}$ kg) | (2.42 x $10^{-17}$ s) | (1.60 x $10^{-19}$ C) | (1 rad) | 1<br>1 | $\eta = k_C = m_e$<br>$= e = \hbar = 1$ |
| Planck | (1.62 x $10^{-35}$ m) | (2.18 × $10^{-8}$ kg) | (5.39 × $10^{-44}$ s) | (1.88 × $10^{-18}$ C) | (1 rad) | 1<br>1 | $\eta = k_C = c$<br>$= G = \hbar = 1$ |

Note 1: Currently within the SI, $k_C = 10^{-7}$ m.kg.$C^{-2}$ x $c^2$ = 8.98 x $10^9$ $m^3$.kg.$s^{-2}$.$C^{-2}$. The numerical factor of $10^{-7}$ between $k_C$ and $c^2$ would no longer be exact in the proposed revised SI, but it will always be very close to this number.

Note 2: The exact conversion factor between coulombs and abcoulombs is $(10^9 k_C)^{1/2}/c$, where $c$ is the value in m/s. Its numerical value would no longer be exactly 10 in the proposed revised SI, but it will always be very close to this number.

Note 3: The exact conversion factor between coulombs and statcoulombs is $(10^9 k_C)^{-1/2}$. Its numerical value would no longer be exactly $1/10c$ (where $c$ is the value in m/s) in the proposed revised SI, but it will always be very close to this number.

Table 1: the size of the five base units in various unit systems, together with any simplifying conventions. Where a base unit is treated as a natural unit, the size of the natural unit is put in brackets. Where base units are linked together this is shown by the equation in the unit box. Numbers are given to three significant figures. The "Underlying SI" is a major modification to the SI within which the radian is treated as a normal (dimensional) base unit. The "SI" row applies to both the current and revised SI. cgs-es, cgs-em, and cgs-Heaviside-Lorentz are variants of the centimetre-gram-second unit systems within which there is no electrical base unit. The "atomic" and "Planck" rows refer to systems using "natural units", as specified in the Features column. $c$, $m_e$, $e$, $\hbar$ and $G$ are the speed of light, the rest mass of the electron, the magnitude of the charge on the electron, the reduced Planck constant and the gravitational constant, respectively. $\eta$ and $k_C$ are defined in the text.

A further convention for electrical and magnetic equations - rationalisation

For the cgs-es and cgs-em unit systems, the choice of whether or not to rationalise the equations of electromagnetism affects the size of the electrical unit, so rationalisation has historically been treated as part of the unit system. However, this is not the case for unsimplified unit systems, where rationalisation can be considered separately.

Rationalisation is in essence the choice to express Gauss's Law (one of the Maxwell equations) as $\nabla \cdot \mathbf{D} = \rho$, where $D$ is the electric displacement field and $\rho$ is the charge density, rather than the unrationalised version $\nabla \cdot \mathbf{D} = 4\pi\rho$. An analogous choice is made for the magnetic field strength $H$, where the rationalised version is $\oint \mathbf{H} \cdot \mathbf{ds} = I$ rather than $\oint \mathbf{H} \cdot \mathbf{ds} = 4\pi I$, $I$ being the current through a loop described by line segments $ds$. For unsimplified unit systems, rationalisation can be done simply by redefining $D$ and $H$ so that the factors relating them to the electric field $E$ and the magnetic flux density $B$, respectively, are changed by a factor $4\pi$. With cgs-es, in contrast, the simplifying constraint means that a reduction in the size of the electrical unit by a factor of $(4\pi)^{1/2}$ is necessary. The process of rationalisation is discussed further in the next section.

As previously noted, it is convenient when rationalising to treat $k_C$ in Coulomb's Law $F = k_C q_1 q_2/r^2$ as $k_C'/4\pi$. The rationalisation decision is conventionally conveyed by writing out this equation in the unit system being used. For the SI this is $F = q_1 q_2/4\pi\varepsilon_0 r^2$, as above. For clarity it is useful also to set out the analogous equation for the force between two current-carrying wires, which for the SI is $F/L = \mu_0 I_1 I_2/2\pi d$, where the meaning of the symbols is given later on.

Changing the subject: the "Gaussian unit system"

The main alternative to the SI for theoretical physicists is known as the Gaussian system of units (Jackson, 1999). How does this fit in to the classification set out above? We suggest that it is already there, and the difference between the Gaussian system and cgs-es as normally used is simply their definition of the magnetic flux density $B$, something which is outside the scope of the unit system. The usual definition of $B$ is $\mathbf{F} = q\mathbf{v} \times \mathbf{B}$, where $F$ is the force on a charge $q$ moving with velocity $v$. In the Gaussian system the definition is $\mathbf{F} = q\mathbf{v} \times \mathbf{B}/c$. The "B"s in these two equations are not the same quantity expressed in different units; they are different quantities with different dimensions: force/(charge x velocity) and force/charge respectively. We suggest that the quantity used in the Gaussian system, which is also used within the Heaviside-Lorentz system, should be called by another name such as "relativistic $B$" and given a different symbol, such as $B_R$, so that $\mathbf{F} = q\mathbf{v} \times \mathbf{B_R}/c$ and $\mathbf{B_R} = c\mathbf{B}$.

The electric field $E$ and the quantity $B_R$ have the same dimensions (force/charge) in any set of units, including SI. Similarly, while the fields $D$ (defined as $\mathbf{D} = \varepsilon_0 \mathbf{E} + \mathbf{P}$) and $H$ (defined as $\mathbf{H} = \mathbf{B}/\mu_0 - \mathbf{M}$) have different dimensions (charge/area and current/length respectively), if we define $\mathbf{H_R} = \mathbf{H}/c = \mathbf{B_R}/c^2\mu_0 - \mathbf{M_R}$, both $D$ and $H_R$ have the dimensions of charge/area. Incidentally, the relativistic magnetic vector potential $A_R$, defined by $\mathbf{B_R} = \nabla \times \mathbf{A_R}$, has the dimensions of voltage, unlike the conventional $A$ which has dimensions of force/current or voltage/velocity.

In the SI, the Maxwell equations relating $E$ and $D$ with $B_R$ and $H_R$ are:

$$\nabla \cdot \mathbf{D} = \rho$$

$$\nabla \cdot \mathbf{B_R} = 0$$

$$\nabla \times \mathbf{E} + \frac{1}{c}\frac{\partial \mathbf{B_R}}{\partial t} = 0$$

$$\nabla \times \mathbf{H_R} = \frac{1}{c}\left(\mathbf{J} + \frac{\partial \mathbf{D}}{\partial t}\right)$$

These are identical in form to those used within the Heaviside-Lorentz system, which of course should also use the symbols $B_R$ and $H_R$ in place of the customary $B$ and $H$.

The last of these equations differs between the two systems when written in terms of $B_R$ rather than $H_R$, where we have:

$$\nabla \times \mathbf{B_R} = \mu_0 c \left(\mathbf{J} + \varepsilon_0 \frac{\partial \mathbf{E}}{\partial t}\right) \qquad \text{(SI)}$$

and

$$\nabla \times \mathbf{B_R} = \frac{1}{c}\left(\mathbf{J} + \frac{\partial \mathbf{E}}{\partial t}\right) \qquad \text{(Heaviside} - \text{Lorentz)}$$

This difference can be seen as a direct result of the simplification convention for Heaviside-Lorentz set out in Table 1, whereby $k_C$ is set equal to $1/4\pi$, making $\varepsilon_0$ equal to 1, and $\mu_0$ equal to $1/c^2$.

The Gaussian unit system is the unrationalised version of the Heaviside-Lorentz system. As mentioned above, the process of rationalising an unsimplified unit system, such as the SI when angles are not involved, is relatively straightforward. Here we just need to write the SI equations in terms of $\mathbf{E}$, $\mathbf{B_R}$, $\mathbf{D_U} = 4\pi\mathbf{D}$ and $\mathbf{H_{RU}} = 4\pi\mathbf{H_R}$, where the subscript U denotes unrationalised, so that:

$$\nabla \cdot \mathbf{D_U} = 4\pi\rho$$

$$\nabla \cdot \mathbf{B_R} = 0$$

$$\nabla \times \mathbf{E} + \frac{1}{c}\frac{\partial \mathbf{B_R}}{\partial t} = 0$$

$$\nabla \times \mathbf{H_{RU}} = \frac{4\pi}{c}\mathbf{J} + \frac{1}{c}\frac{\partial \mathbf{D_U}}{\partial t}$$

These SI equations are identical in form to the Gaussian version of the Maxwell equations. The advantage of the Gaussian and Heaviside-Lorentz systems for theoretical physicists in terms of magnetic and electric fields having the same units, as is natural in special relativity, can be seen to be readily available within the SI, simply by paying more careful attention to how the quantities are defined.

<u>Suggested modifications to the summary description of the revised SI</u>

The proposed summary definition of the revised SI (BIPM, draft 9[th] edition of the SI Brochure, 2016) is:

"The International System of Units, the SI, is the system of units in which

- the unperturbed ground state hyperfine transition frequency of the caesium 133 atom $\Delta \nu_{Cs}$ is 9 192 631 770 Hz,
- the speed of light in vacuum $c$ is 299 792 458 m/s,
- the Planck constant $h$ is 6.626 070 0XX ×10$^{-34}$ J s,
- the elementary charge $e$ is 1.602 176 62X X ×10$^{-19}$ C,
- the Boltzmann constant $k$ is 1.380 648 XX ×10$^{-23}$ J/K,

- the Avogadro constant $N_A$ is 6.022 140 8XX ×$10^{23}$ $mol^{-1}$,
- the luminous efficacy $K_{cd}$ of monochromatic radiation of frequency 540 ×$10^{12}$ hertz is 683 lm/W,"

The inclusion of "XX" in the values of the newly defined constants denotes that they are not yet finally agreed. This statement defines the sizes of the SI base units, albeit in an indirect way, but does not cover the conventions about $\varepsilon_0$, $\mu_0$, and angle, which are required for a full description of the unit system as it is to be practically used.

We suggest that the text should therefore be supplemented by:

"and in which

- the equation describing the force between two electric charges in a vacuum is given by $F = q_1 q_2 / 4\pi\varepsilon_0 r^2$, where $F$ is force expressed in units of N, $q$ is electrical charge in C, $r$ is the separation in m, and $\varepsilon_0$ is the permittivity of free space in units $m^{-3}\ kg^{-1}\ s^4\ A^2$,

- the equation describing the force between two infinite, parallel current-carrying wires of negligible cross-section, in a vacuum, is given by $F/L = \mu_0 I_1 I_2 / 2\pi d$, where $F/L$ is the force per unit length expressed in units of N/m, $I$ is electrical current in A, $d$ is the separation in m, and $\mu_0$ is the permeability of free space in units $m\ kg\ s^{-2}\ A^{-2}$,

- the equation describing the arc length $s$ of a sector of radius $r$ and angle $\theta$ is given by $s = r\ \theta$, where $s$ and $r$ are in m and $\theta$ is in rad,

- the equation describing the area $A$ on the surface of a sphere of radius $r$ defined by a solid angle $\Omega$ at its centre is given by $A = \Omega\ r^2$, where $A$ is in $m^2$, $r$ is in m and $\Omega$ is in sr."

All of these points apply equally to the current SI and the revised SI. The first two points are covered by the (current) 8th Edition of the SI Brochure (in Section 1.2), but are omitted from the draft 9th Edition. The latter two points, which have the feature of both specifying the equations to be used and defining the units radian and steradian, do not appear in either the 8th Edition or the draft 9th Edition.

Although we have treated the radian as a dimensional base unit within the unsimplified unit system we have called the "underlying SI", we do not advocate a change in the status of angle within the SI itself.

<u>Implications for conversion between unit systems, dimensional analysis and software</u>

Although the simplified unit systems described in this paper all have their uses, we suggest that many activities are best carried out with the simplifying conventions removed, so that each base quantity has an independent dimensional base unit with a well-defined value. For example:

- Conversion between unit systems. Any quantity can be expressed in a different set of units by following a simple process. This involves expressing the quantity in base units and converting each base unit by the appropriate factor. This is only possible if the values of the base units in the different systems are available. It should be stressed that conversion of quantities between unit systems can only be done correctly when the quantities are clearly defined. This is particularly important for the magnetic quantities given the symbols $B$ and $H$;

- Dimensional analysis. As implied earlier, the convention of linking base units in order to simplify equations, as in the cgs-em and cgs-es systems, should not be seen as changing the physical dimensions of the quantities involved. For the purposes of dimensional analysis, the dimensions of a quantity are best seen as those described by the unsimplified system of units. For example, in cgs-em, electrical resistance can be given the same units as velocity, but this does not mean that the two quantities should be seen as having the same dimensions;

- Software. As a general rule, procedures which involve conventions that need to be understood by the user are difficult to accommodate within software. Many such conventions can be avoided by using an unsimplified set of units, such as the "underlying SI", within the software. The disadvantage of equations containing a larger number of constants is much less serious for software than for manual calculation, and is offset by the advantage of the unit algebra being completely transparent.

## Conclusions

Unlike the laws of physics, the unit systems in which they are expressed are fundamentally pragmatic creations. Scientists and mathematicians have an understandable tendency to use conventions within unit systems to simplify equations. The result of this is a blurring of the formal categories "base" and "derived" units, where the base units are ideally both independent and considered to have orthogonal dimensions. This blurring occurs for electromagnetic quantities in the case of cgs systems, for angle in the case of all familiar unit systems including the SI, and for all quantities when natural units are used.

We conclude that unit systems should only be specified in terms of their "base" units alone when no simplifying conventions apply, and when this is not the case, the conventions should be made explicit. The unsimplified version of any unit system is particularly useful in the circumstances described in the last section, and should not be considered to have been made redundant because a simplified version of it is in common use.

We also conclude that the radian's status within the SI is not well described by the term dimensionless derived unit, as is currently the case. The previous term supplementary unit would be preferable. The description of the radian as "an SI unit that is defined independently of the SI base units, which is by convention treated as dimensionless because this simplifies equations" would be more correct, but this would be too complicated for frequent use. Such a change in terminology would have no effect on the radian's practical use, or on equations which contain angles, but would bring greater clarity to a longstanding issue.

## Acknowledgements

We would like to thank Richard Davis for stimulating discussions, and helpful comments on the manuscript.

## References

BIPM *Le Système International d'Unités (SI)* 8[th] edn (Sèvres: Bureau International des Poids et Mesures) (2006)

BIPM *Le Système International d'Unités (SI)* Draft 9[th] edn (Sèvres: Bureau International des Poids et Mesures) (2016), available through http://www.bipm.org/en/measurement-units/new-si/#communication

J B Brinsmade, Plane and Solid Angles; their Pedagogic Value When Introduced Explicitly, American Journal of Physics **4**, 175-179 (1936)

K R Brownstein, Angles – Let's treat them squarely, American Journal of Physics **65**(7) 605-614 (1997)

J de Boer, On the History of Quantity Calculus and the International System, Metrologia, 31, Number 6, 405-429 1995

M Duff, How fundamental are fundamental constants? Contemporary Physics, 56, 35-47 (2015)

F Frezza, S Maddio, G Pelosi, and S Selleri, The Life and Work of Giovanni Giorgi: The Rationalization of the International System of Units, IEEE Antennas & Propagation Magazine, 152-165, December 2015.

J D Jackson J D, Classical Electrodynamics, 3$^{rd}$ Ed, John Wiley (1999)

P J Mohr and W D Phillips, Dimensionless units in the SI, Metrologia **52** 40–7 (2014)

P Quincey and R J C Brown, Implications of adopting plane angle as a base quantity in the SI, Metrologia **53** 998-1002 (2016)

P Quincey, Natural units in physics, and the curious case of the radian, *Phys. Educ.* **51** 065012 (2016)

John Roche, The mathematics of measurement, a critical history, The Athlone Press (1998)

A B Torrens, On angles and angular quantities, Metrologia **22** 1-7 (1986)

Wilczek F, Fundamental constants, arXiv:0708.4361v1 (2007)